\documentclass{elsart3}
\usepackage{graphicx}
\begin{document}

\begin{frontmatter}
%%%%%%%%%%%%%%%%%%%%%%%%%%%%%%%%%%%%%%%%%%
%%%% Title here
%%%%%%%%%%%%%%%%%%%%%%%%%%%%%%%%%%%%%%%%%%
\title{Twin spin/charge roton mode and superfluid density:  
primary determining factors of $T_{c}$ in high-$T_{c}$ superconductors
observed by neutron, ARPES, and $\mu$SR}
%%%%%%%%%%%%%%%%%%%%%%%%%%%%%%%%%%%%%%%%%%
%%%% List of authors  (edit the following)
%%%%%%%%%%%%%%%%%%%%%%%%%%%%%%%%%%%%%%%%%%
\author{Y.~J.~Uemura}
\corauth[cor]{Tel. +1 212 854 8370,
Fax: +1 212 854 3379,
email: tomo@lorentz.phys.columbia.edu} 
%\author[OXF]{B. Ware}
%\author[RAL]{X. Pedient}
%\author[LIV]{M.P. Three} 
%%%%%%%%%%%%%%%%%%%%%%%%%%%%%%%%%%%%%%%%%%
%%%% List of addresses (edit the following)
%%%%%%%%%%%%%%%%%%%%%%%%%%%%%%%%%%%%%%%%%%
\address{Physics Department, Columbia University, 538 West, 120th Street,
New York, NY 10027, USA}

%%%%%%%%%%%%%%%%%%%%%%%%%%%%%%%%%%%%%%%%%%
%%%% Abstract
%%%%%%%%%%%%%%%%%%%%%%%%%%%%%%%%%%%%%%%%%%
\begin{abstract}
{In the quest for primary factors which determine the 
transition temperature $T_{c}$ of high-$T_{c}$ cuprate
superconductors (HTSC), we develop a phenomenological picture
combining experimental results from muon spin relaxation 
($\mu$SR), neutron and Raman scattering, and angle-resolved
photoemission (ARPES)
measurements, guided by an analogy with superfluid $^{4}$He.
The 41 meV neutron resonance mode and the ARPES 
superconducting coherence peak (SCP) can be viewed
as direct observations of spin and charge soft modes,
respectively, appearing near ($\pi,\pi$) and the center
of the Brillouin zone, having identical energy transfers
and dispersion relations.  We present a conjecture that the
mode energy of this
twin spin/charge collective excitation, as a roton analogue in HTSC,
plays a primary role in determining $T_{c}$, together with the superfluid density
$n_{s}/m^{*}$ at $T \rightarrow 0$.  
We further propose a microscopic model for pairing based on a resonant spin-charge 
motion, which explains the extremely strong spin-charge coupling, relevant
energy scales, disappearence of pairing in the overdoped region, and 
the contrasting spin-sensitivities of nodal and antinodal charges in HTSC systems.
Comparing collective versus single-particle excitations, pair formation versus
condensation, and local versus long-range phase coherence,
we argue that many fundamental features of HTSC systems, including
the region of the Nernst effect, can be
understood in terms of condensation and fluctuation phenomena of bosonic correlations 
formed above $T_{c}$.}
\end{abstract}

%%%%%%%%%%%%%%%%%%%%%%%%%%%%%%%%%%%%%%%%%%
%%%% Keywords
%%%%%%%%%%%%%%%%%%%%%%%%%%%%%%%%%%%%%%%%%%
\begin{keyword}
high-$T_{c}$ superconductivity, $\mu$SR, ARPES, neutron scattering.
\end{keyword}
\end{frontmatter}

\section{Introduction}
Since the author joined the research group of Prof. Toshimitsu Yamazaki 
at the Univ. of Tokyo in 1976, the ultimate goal of our research has been to 
combine powerful modern experimental techniques with insightful physics
phenomenology.  In this lecture, we would like to present our attempt
towards this goal in 18 years of study on high-$T_{c}$ cuprate superconductors (HTSC),
motivated by the quest for primary factors which determine $T_{c}$,
and guided by an instructive analogy between HTSC and superfulid
He systems.  Some seed ideas of the present work have been discussed in ref. [1].
%\begin{figure*}[htb]
%\centering
%% uncomment following line when using a real figure
%%\includegraphics[width=\columnwidth]{figure1.ps}
%\includegraphics[width=3in,angle=270]{fig8land.eps}
%\caption{Figure caption.
%}
%\label{fig1}
%\end{figure*}
%%\end{figure}

\begin{figure}[htb]
\centering
% uncomment following line when using a real figure
\includegraphics[width=\columnwidth]{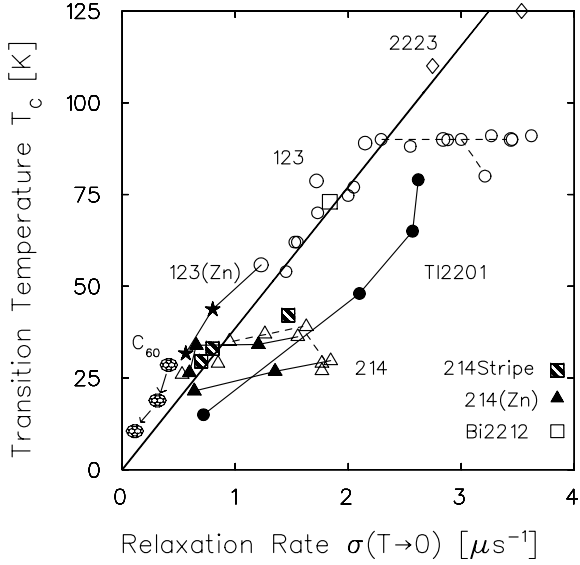}
\caption{Muon spin relaxation rate $\sigma\propto n_{s}/m^{*}$ at $T\rightarrow 0$ 
from various high-$T_{c}$ cuprate superconductors [2,1] and 
A$_{3}$C$_{60}$ systems plotted against the superconducting
transition temperature $T_{c}$.  The points for HTSC with open symbols
represent simple hole doped systems, while closed triangles
are for (Cu,Zn) substitution [3], ``stripe'' symbols for
systems with formation of island regions with incommensurate
static spin modulations [4], and closed circles for overdoped Tl2201 [5].}
\label{fig1}
\end{figure}

\section{Superfluid Energy Scales vs. $T_{c}$}

By 1989, within a few years after the discovery of 
HTSC systems, we established [1,2] strong correlations between 
$T_{c}$ and $n_{s}/m^{*}$ (superconducting carrier density /
effective mass) at $T\rightarrow 0$ by measuring 
the muon spin relaxation rate $\sigma$ and the London penetration 
depth $\lambda$ as $\sigma \propto 1/\lambda^{2} \propto n_{s}/m^{*}$.
As shown in Fig. 1, a nearly linear relationship between $T_{c}$ and $n_{s}/m^{*}$
(often referred to as the ``superfluid density'') holds in 
most of the underdoped cuprates as well as in HTSC systems with Cu/Zn substitutions [3],
formation of island regions having static stripe spin correlations [4],
and overdoped systems having microscopic phase separation between superfluid and normal
unpaired fermions [5,6].  The strong correlation
was one of the earliest signatures suggesting fundamental difference of HTSC
systems from BCS superconductors.  The robustness of this relationship
against heterogeneity, analogous to the case in superfluid He films 
and $^{3}$He/$^{4}$He mixture films on regular and porous media [1], can be expected only
for systems with very strong coupling.  In Fig. 1, we also notice that
the 214 systems exhibit an ``early branching off'' from the nearly linear
relationship, which we will ascribe to the closeness to competing magnetic
states, as discussed later.

\begin{figure}[t]
\centering
% uncomment following line when using a real figure
\includegraphics[angle=90, width=\columnwidth]{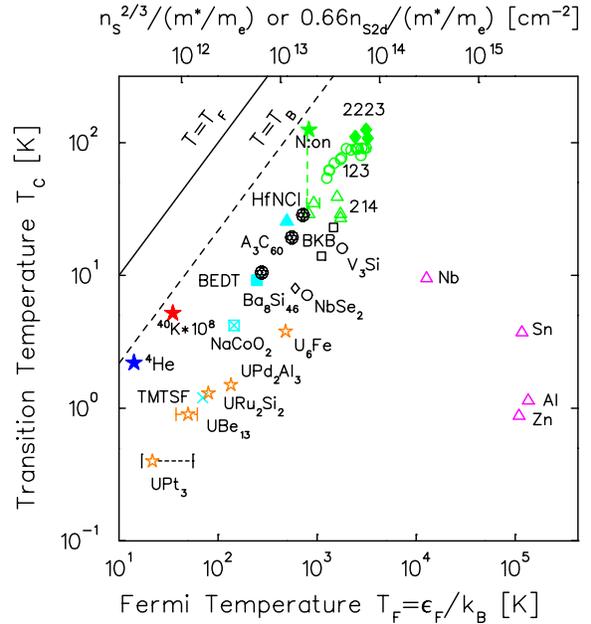}
\caption{
(color) Plot of $T_{c}$ versus the effective Fermi temperature
$T_{F}$ obtained from the superfluid response $n_{s}/m^{*}$
of various superconducting systems, 
first attempted in ref. [7] in 1991, and updated [1].  
We see an empirical upperlimit $T_{c}/T_{F} \sim 0.05$ 
for superconducting systems.  Also included are the corresponding
points for the superfluid $^{4}$He (blue star).  The $T_{B}$ line shows the 
BE condensation temperature for the 
ideal bose gas of boson density $n_{s}/2$ and 
mass $2m^{*}$.  The green star represents the onset 
temperature $T_{on}$ of the 
Nernst effect, shown in Fig. 12, for La$_{1.9}$Sr$_{0.1}$CuO$_{4}$ [29].}
\label{fig2}
\end{figure}

In 1991 [7], we converted the superfluid density into the effective Fermi
temperature $T_{F}$, which represents 
an effective kinetic energy scale of mobile superconducting carriers
corresponding to the ``Drude'' spectral weight in optical responses.
The results in Fig. 2 are shown together with the Bose-Einstein (BE) condensation
temperature $T_{B}$ expected when all the superconducting carriers form a non-interacting
ideal Bose gas with the density $n_{s}/2$ and mass 2$m^{*}$.  The transition 
temperatures $T_{c}$ of HTSC, organic BEDT, and some other novel superconductors are reduced
by a factor 4-5 from $T_{B}$, while the above-mentioned linear relationship
emerges to be parallel to the $T_{B}$ line.  This feature suggests a fundamental importance of 
the BE condensation concept in cuprates.  The blue star point represents the superfluid
transition of (3-dimensional) liquid He, at $T_{\lambda}$ = 2.2 K
which is about 50 \%\ reduced from the $T_{B} = 3.2$ K calculated with the number density
and mass.  Figure 2 can be viewed as an experimentalist's way to classify 
various superconductors and superfluids between the two extrema of idealized BE and BCS 
condensations in the strong and weak-coupling limits. 

\section {BE-BCS crossover and phase diagram of HTSC}

\begin{figure}[t]
\centering
% uncomment following line when using a real figure
\includegraphics[width=\columnwidth]{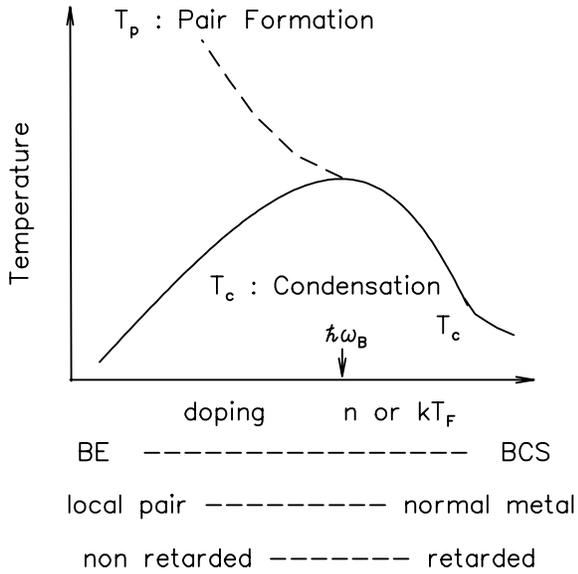}
\caption
{The BE-BCS crossover picture proposed by Uemura [8,9] in 1994 with the 
crossover region characterized by the matching of kinetic energy
$k_{B}T_{F}$ of the condensing carriers with the mediating boson
energy $\hbar\omega_{B}$.  When one identifies the pair formation 
temperature $T_{p}$
as the pseudo-gap temperature $T^{*}$, this phase diagram can be
mapped to the case of HTSC.}
\label{fig3}
\end{figure}

In 1993-94 we proposed a conjecture [8-10] that the phase diagram of HTSC systems can be
mapped onto a general concept of BE to BCS crossover shown in Fig. 3.  
When bosonic correlations develop well above $T_{c}$,
BE condensation occurs when the
thermal wave length of bosons becomes comparable to the interboson distance, 
resulting in $T_{c} \propto n_{s}/m^{*}$ in 2-d and
$n_{s}^{2/3}/m^{*}$ in 3-d systems.
In the BCS condensation, the attractive ``pairing'' energy scale 
(i.e., the gap) determines $T_{c}$, since the 
number-density restriction for BE condensation is readily satisfied when the 
Cooper pairs (bosons)
are formed at $T_{c}$.  The $T_{c}$ v.s. $n_{s}/m^{*}$ correlations and the pseudogap
behavior lead to a view that the underdoped cuprates
are in the BE region, with the $T^{*}$ representing pair (boson)
formation above $T_{c}$.  In this mapping, the optimal-doping region 
lies in the crossover region, where the number-density
energy scale (the effective Fermi temperature $T_{F}$) is close to the 
energy of pair mediating interaction $\hbar\omega_{B}$.  
In HTSC systems, the optimum
region appears around $T_{F} \sim$ 2000 K.  
Hence the antiferromagnetic spin fluctuations, having a similar
energy scale, become an attractive candidate for the pairing mediator [10].

\begin{figure}[t]
\centering
% uncomment following line when using a real figure
\includegraphics[width=\columnwidth]{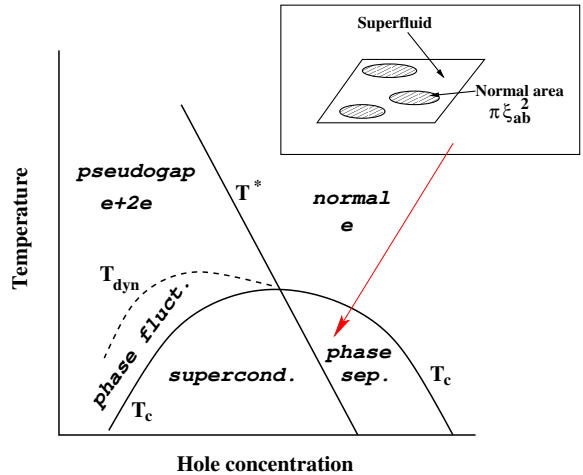}
\caption
{(color) A generic phase diagram proposed for cuprates by 
Uemura [1,12], including the distinction between the 
pair formation $T^{*}$ and the onset temperature of 
dynamic superconductivity $T_{dyn}$ which corresponds to the 
$T_{on}$ of the Nernst effect in Fig. 12.  Dynamic superconductivity
is associated with phase fluctuations of bosonic wavefunctions.
The inset illustrates the microscopic phase separation
between superconducting and normal metal regions in the 
overdoped region [6].}
\label{fig4}
\end{figure}

Figure 4 shows our current view of the actual HTSC phase diagram [1,12].
Several experimental results [5-6] in the overdoped region indicate existence 
of microscopic phase separation, as we first suggested in 1993 [5] and 
as has been directly confirmed very recently by scanning tunneling microscopy (STM) 
studies [11].  
The overdoped region is thus quite different from the simple BCS situation.
When the pairing strength $T^{*}$ rapidly decreases 
with increasing doping (increasing $T_{F}$), the system responds by 
creating a ``phantom superconducting'' region where superconductivity
is maintained by microscopic phase separation to gain condensation energy
at the expense of Coulomb energy for charge disproportionation [6].

\section {Collective soft-mode energies determining $T_{c}$}

\begin{figure}[htb]
\centering
% uncomment following line when using a real figure
\includegraphics[width=\columnwidth]{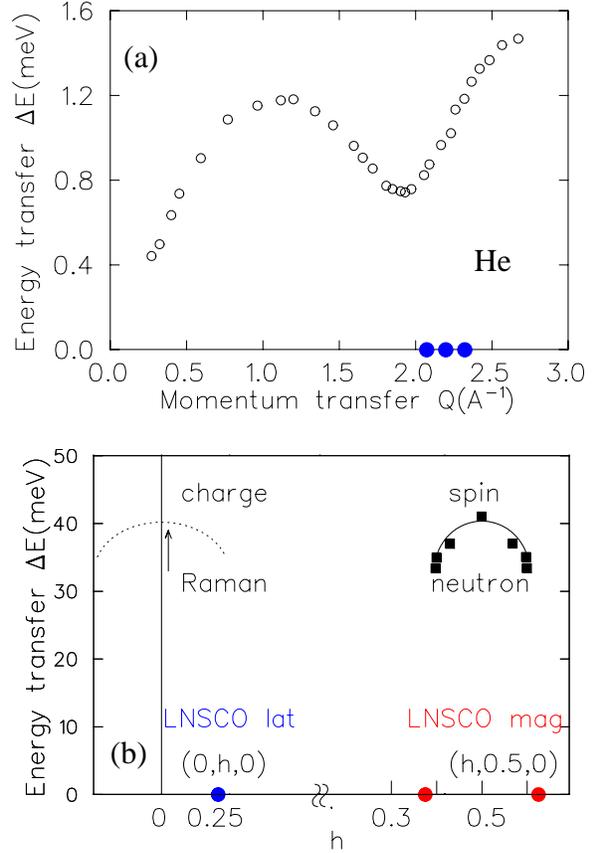}
\caption
{(color) (a) The dispersion relation of phonon-roton excitations in superfluid
$^{4}$He observed by neutron scattering [13].  
The closed circles show the Bragg points of HCP solid He [14]
at the pressure of 66 atmospheres at T = 1.1 K.
(b) The dispersion relation around the ($\pi,\pi$) resonance mode
observed in YBCO by neutrons [16] (closed squares), 
the location of the satellite Bragg peaks (red =  magnetic; blue = lattice, estimated
from the adjacent Brillouin zone)
found in the static spin/charge stripe system 
(La,Nd,Sr)$_{2}$CuO$_{4}$ (LNSCO) [18],
and the proposed charge branch of the twin spin/charge roton
(dotted line), to which the Raman A1$_{g}$ response and the ARPES SCP in Figs. 6 and 8 are ascribed.}
\label{fig5}
\end{figure}

In superfluid 3-d liquid $^{4}$He, the reduction of the 
lambda temperature $T_{\lambda}$ = 2.2 K from the idealized $T_{B}$ = 3.2 K
is caused by the existence of roton excitation, which ia a 
soft phonon mode towards solidification of He.  Figure 5(a) shows
the phonon-roton dispersion relation [13] together with the Bragg
peaks of solid He [14].  The effect of roton on $T_{c}$ can be clearly seen via the 
nearly linear relationship in Fig. 6 which plots
$T_\lambda = T_{c}$ versus the roton mode energy (observed at different
pressures [15]: orange-star symbols), where the He values for the both 
axes are multiplied by a factor 60 to be compared with the cuprates. 
Helium rotons demonstrate that the closeness to the competing state (solid He) 
influences the condensation temperature of a bosonic superfluid.

\begin{figure}[htb]
\centering
% uncomment following line when using a real figure
\includegraphics[width=\columnwidth]{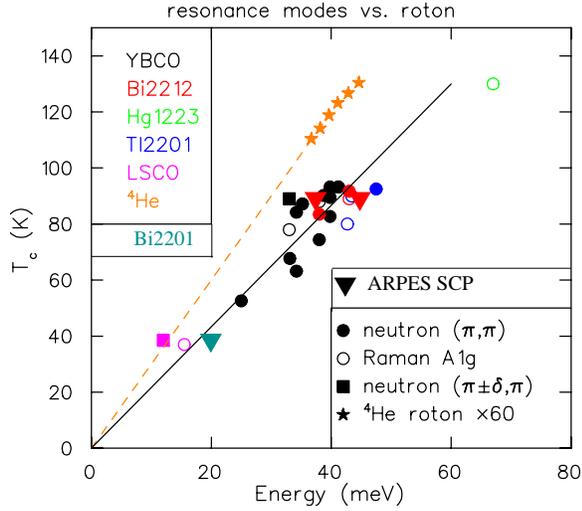}
\caption
{(color) The plot of $T_{c}$ versus energy of the roton minimum of 
bulk superfluid $^{4}$He (values for both axes multiplied by 
a factor 60) measured under applied pressures [15],
compared with the relationship seen in 
HTSC systems for the energies of the neutron resonance mode at 
($\pi,\pi$) [16,1], the Raman A1$_{g}$ mode [17,1],
and the ARPES SCP [23]. 
Also included are the neutron energy transfers of spectral 
weight maximum near the 
($\pi\pm\delta,\pi$) point in YBCO [16] and LSCO [19] (closed squares).}
\label{fig6}
\end{figure}

Guided by an analogy to He, we plot the energy of the 41 meV magnetic
resonance mode of HTSC [16] versus $T_{c}$ in Fig. 6.  Also plotted is
the energy of $A_{1g}$ Raman mode which follows the same relationship [17].
The magnetic resonance mode is a soft-magnon mode towards the magnetically ordered
stripe state competing with the superconducting state, analogous to He rotons
being a soft phonon mode towards a competing state.
The neutron resonance detects an S=1 excitation with the wavevector 
transfer ($\pi,\pi$), while the Raman mode represents an S=0 excitation near the 
zone center.  In most of known cases, the ``2-magnon'' mode in Raman measurements,
having twice the magnon energy, is required to match the nearly zero net momentum transfer
in Raman measurements.
In contrast, Fig. 6 shows a one-to-one correspondence of neutron and Raman energies.
This apparent contradiction can be resolved by assuming a roton-analogue twin
mode in HTSC [1], having the spin-branch near ($\pi,\pi$) and charge branch near
the zone center, with the same energy minimum and dispersion, as illustrated
in Fig. 5(b).  

Indeed the competing states in the cuprates, such as the stripe state
or the Mott insulator antiferromagnetic state, have both spin and charge
orders, with the spin Bragg peak (red circles in Fig. 5(b)) and charge
Bragg peak (blue circles) appearing when that state wins against the 
superconducting state [18].
When the superconducting state wins, the spin-charge modulations become
dynamic, yet conserving the very strong correlations with each other and
having the same soft-mode energy transfers corresponding to the 
``superconducting condensation energy''.  Via this mechanism, although neutrons and 
photons (in the Raman studies) create different final states (S=1 and S=0 states)
with different momentum transfers, the spin and charge branches appear with 
the same energy transfers.  The magnon soft-mode can exist only when it is associated with
the partner dynamic charge modulation.  Thus the excitation of the 41 meV mode contributes to 
the reduction of superfluid density, playing a role similar to He rotons, leading
to a linear relationship between $T_{c}$ and the mode energy for HTSC systems
as shown in Fig. 6.  This analogy between He and HTSC is strengthened by their 
nearly equal ratios of mode energy and $T_{c}$. 

\section {Common spin-charge dispersion relation and ARPES coherence peak as the charge branch}

\begin{figure}[t]
\centering
% uncomment following line when using a real figure
\includegraphics[width=\columnwidth]{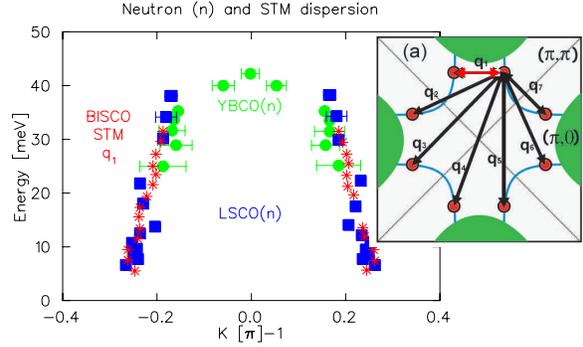}
\caption
{(color) The dispersion relation of the 
neutron resonance mode, measured around the 
momentum transfer ($\pi,\pi$) in YBCO and 
LSCO [19] systems near optimal doping, compared with 
the dispersion of the $q_{1}$ charge scattering mode
measured by the Fourier transform STM spectroscopy
in Bi2212 [21]. The momentum transfer for the latter is scaled by a factor 2
to account for the difference between the spin modulation with the 
1/8 periodicity and the charge modulation with the 1/4 periodicity,
connecting the Fermi arcs near the antinodal points
($0,\pi$). The inset figure shows the Brillouin zone
and the STM $q_{1}$ mode with the red arrow.}
\label{fig7}
\end{figure}

\begin{figure*}[t]
\centering
%% uncomment following line when using a real figure
%%\includegraphics[width=\columnwidth]{figure1.ps}
\includegraphics[width=6in]{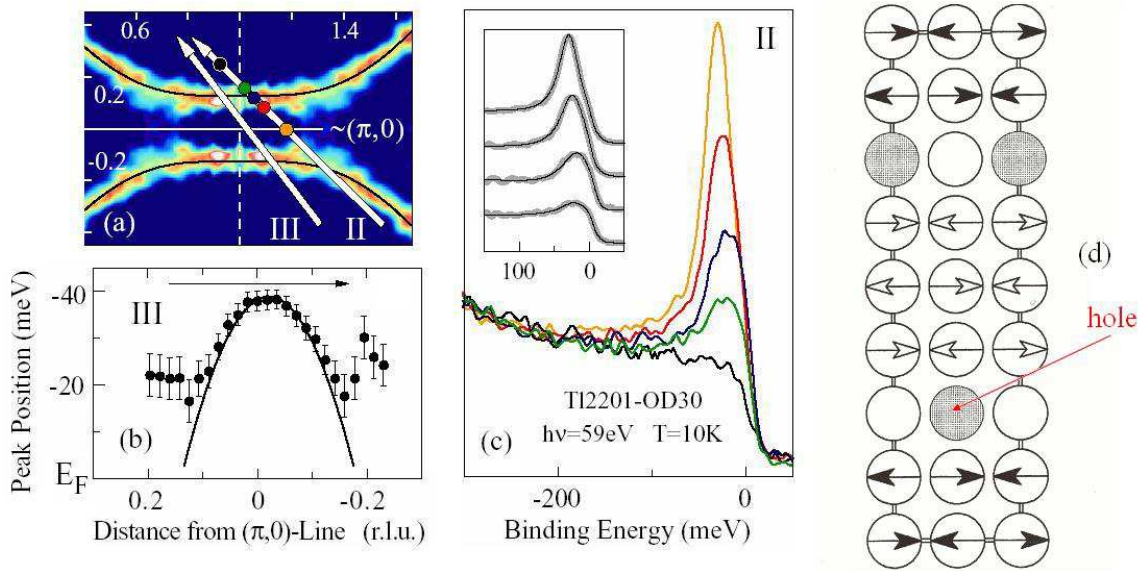}
\caption
{(color) The ARPES superconducting coherence peak (SCP) observed in overdoped 
Tl$_{2}$Ba$_{2}$CuO$_{6+\delta}$ (Tl2201) with $T_{c}$ = 30 K [22].  (a) The 
trajectory of momentum in the reciprocal space near the antinodal point
($\pi,0$).  (b) The dispersion relation observed along the line III in (a), 
which is nearly identical to those
shown in Fig. 7.  Note that the peak position energy, multiplied by (-1),
corresponds to the energy input given by the ARPES photon to the system.  
(c) The intensity versus energy profile of the SCP measured along the line II in (a).
(d) Real-space pattern of the stripe spin-charge modulations in the cuprates
[18].  Doped holes reside with Cu atoms shown by the shaded area, which have
neighbouring Cu spins with opposite directions.  This leads to 
the ``geometrical frustration'' for the 
spin of a charge liberated in ARPES from 
the dynamic stripe spin-charge correlations.} 
\label{fig8}
\end{figure*}
%%\end{figure}

The 41 meV spin-branch exhibits a dispersion relation, extending downwards towards the 
satellite stripe Bragg peaks, as shown for YBa$_{2}$Cu$_{3}$O$_{7-\delta}$ (YBCO) and  
La$_{2-x}$Sr$_{x}$CuO$_{4}$ (LSCO) [19] in Fig. 7
and even in non-superconducting La$_{2-x}$Ba$_{x}$CuO$_{4}$
(LBCO) with $x$ = 1/8 [20].  There seem to be two cases: (a) the 
mode energy at ($\pi,\pi$) is reduced 
for systems with lower $T_{c}$'s, such as in the underdoped YBCO; and (b) the dispersion 
itself does not change but the lower-energy branch is populated for systems with lower
$T_{c}$'s, such as in LSCO and LBCO 214 cuprates. When we plot the energy of the
spectral-weight maximum of the lower-energy 
populated branch, which is often referred to as the ``spin gap energy'' in 
neutron studies, the correspondence with superfluid He becomes even better, as shown by
the filled square symbols in Fig. 6.  
Christesen {\it et al.\/} [19] noticed 
that, after a factor 2 adjustment of momentum
transfers, the same dispersion relation is also seen in the $q_{1}$  
charge scattering observed in the STM studies 
in Bi$_{2}$Sr$_{2}$CaCu$_{2}$O$_{8+\delta}$  (Bi2212) [21], shown by the red star
symbols in Fig. 7.

The present author has recently noticed that 
this ``common dispersion of spin and charge responses''
appears also in the ARPES results of the ``superconducting coherence peak'' (SCP),
obtained by Plat\'e {\it et al.\/} [22] in Tl2201 near the antinodal
wavevector ($\pi,0$).  Figure 8 shows the energy-transfer by photons to the 
system in ARPES (i.e., negative peak energy) plotted versus the distance of the momentum
transfer from ($\pi,0$), showing dispersion nearly identical to those in Fig. 7.  Furthermore, 
the energy transfer of the ARPES SCP [23] closely follows the 
relationship found for neutron and Raman modes, as
shown in Fig. 6 with filled-triangle symbols.
These observations suggest that the ARPES SCP may be 
interpreted as a direct manifestation of the charge branch of the 
twin spin-charge soft mode.

\section {Collective mode versus single-particle pair-breaking excitations}  

Although fundamental differences of HTSC from traditional BCS superconductors
were noticed by the mid 1990's, the destruction of superfluid density has almost always been 
discussed exclusively in terms of single-particle pair-breaking excitations:
people's minds have still been controlled by the BCS theory.
This tendency has perhaps been enhanced by the successful explanation of 
the low-temperature penetration depth results by the d-wave nodal quasiparticle excitations.

Single-particle excitations would cost the gap 
energy $\Delta(k)$ in pair breaking.  Imagine 
a spin-singlet charge pair formed to gain the antiferromagnetic exchange energy $J$.
Input of energy $J$ is required to flip the spin in the single-particle process.  
In collective excitations based on many-body correlations, a spin triplet
excitation can be realized as a spin wave which can be obtained without an energy
input of $J$.  This is why the S=1 neutron resonance mode does not cost any pair-breaking energy.
%In other words, such a process would take the ``2e'' superconducting charge
%out of condensate and put that into an excited state of the competing 
%stripe state as ``2e'', without pair breaking.  
Indeed, the mode energy in the 
underdoped HTSC, proportional to $T_{c}$, 
follows the doping dependence completely opposite to 
the behavior of the gap (and pairing) energy $\Delta$ (and $T^{*}$),
as illustrated in Fig. 9 [16].
It is very difficult to ascribe $T_{c}$ to the nodal quasiparticle
pair-breaking excitations [24], in view of this opposite behavior of $\Delta$ and $T_{c}$.

\begin{figure}[t]
\centering
% uncomment following line when using a real figure
\includegraphics[width=\columnwidth]{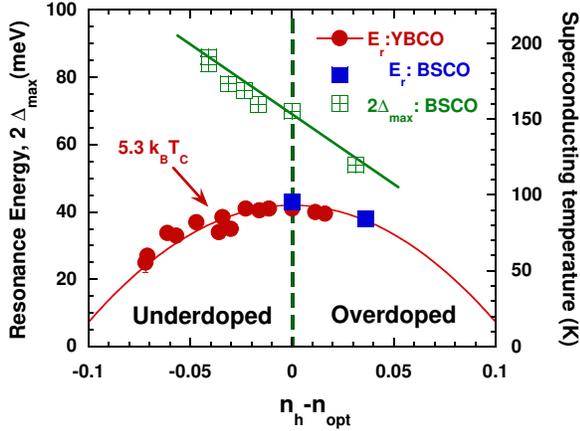}
\caption
{(color) Doping dependence of the resonance mode energy
determined by neutron scattering [16] compared to
the gap energy $2\Delta$ determined by 
tunneling measurements in HTSC systems.  The mode energy scales
with $T_{c}$ in the whole doping range, but the 
gap shows the doping dependence opposite to that
of $T_{c}$ in the underdoped region.} 
\label{fig9}
\end{figure}

From a single-particle type view, an ARPES excitation near the antinodal
($\pi,0$) point requires to break the Cooper pair, costing the 
gap energy $\Delta$, in addition to any bosonic mode energy $\Omega$
to which the charge is coupled.  So, the SCP energy has been discussed
in terms of $\Delta + \Omega$ [25].
However, imagine the case when the energy input by ARPES photon is used
to create collective dynamic spin-charge modulations, as illustrated 
in Fig. 8(d) [18].  Thanks to the microscopic separation of spins and charges
in this stripe pattern, the spin of a hole is ``frustrated'' with the 
neighbouring Cu moments.   This ``geometrical frustration'' makes
the spin direction of the charge irrelevant when the charge is 
knocked out as the ARPES photo-electron:
the charge liberation does not cost the pairing energy $J$ or $\Delta$.
Therefore, we would expect direct correspondence of the ARPES SCP energy
to the boson mode energy $\Omega$ for the case of collective excitations of the
charge soft mode.  This explains the agreement of 
neutron and SCP energies in Fig. 6.  Such a liberated charge would carry momentum 
of the ``broken'' charge-density-wave pair correlation, as clearly seen from 
the ARPES momentum in Fig. 8(a) [22].        

\section {Microscopic pairing via resonant spin-charge motion}

These considerations suggest that pairing in HTSC should involve:
(1) a very strong spin and charge coupling; (2) energy scale
of $T_{F}$ (charge kinetic energy) comparable to that of 
pair-mediation interaction; and (3) disappearance of pairing in the overdoped region.
Guided by these, the present author proposed a pairing mechanism
in 2004 [1] based on resonant spin-charge motion of antinodal charges.

\begin{figure}[htb]
\centering
% uncomment following line when using a real figure
\includegraphics[width=\columnwidth]{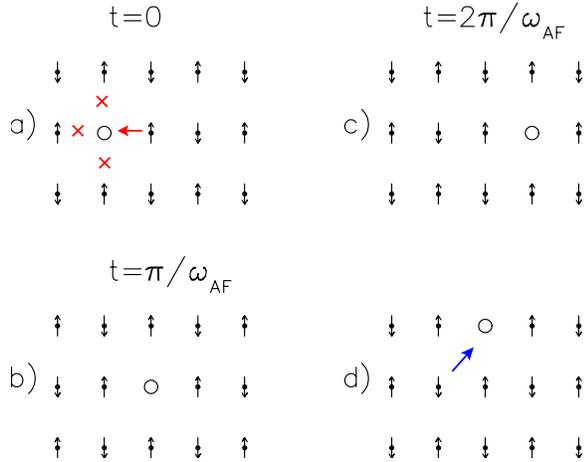}
\caption
{(color) 
An illustration of a charge motion in the cuprate
resonant with the antiferromagnetic 
spin fluctuations with the frequency $\omega_{AF}$.
(a) shows the $t =0$ configuration, and the three frustrated bonds
created when the charge hops while the surrounding spins keep their
directions.
(b) shows the case in which a charge hops 
at $t=\pi/\omega_{AF}$ after a half period when surrounding spins have
changed their directions.
(c) shows the situation after the full period
$t = 2\pi/\omega_{AF}$ for the propagation towards
the ($\pi,0$) direction, representing the case for 
an antinodal charge with wavevector $k = 2\pi/2a$
having $kT_{F} = \hbar\omega_{AF}$.  Spin frustration can be avoided via the 
dynamical motion in (b) and (c).
(d) shows that the diagonal motion of a nodal charge on the same
spin sublattice does not create frustrated
bonds regardless of the time sequence of motion with respect to the 
AF spin fluctuations.}
\label{fig10}
\end{figure}

\begin{figure}[b]
\centering
%% uncomment following line when using a real figure
%%\includegraphics[width=\columnwidth]{figure1.ps}
\includegraphics[width=\columnwidth]{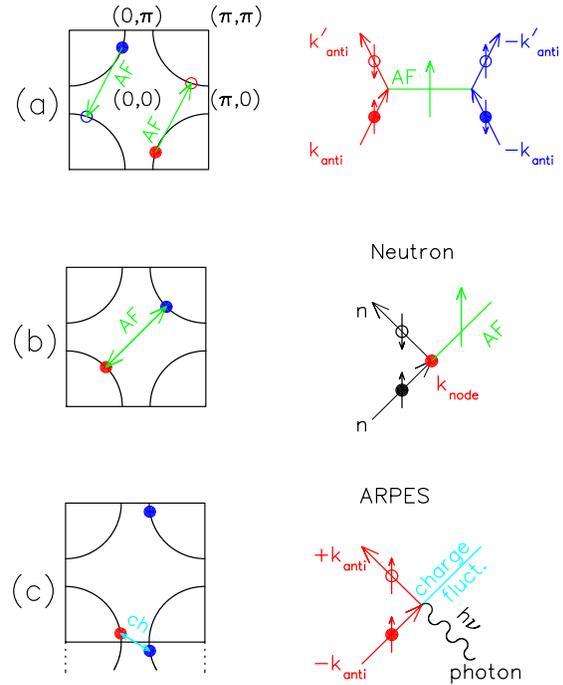}
\caption
{(color) (a)
An illustration of the attractive interaction 
obtained in the scattering process of 
AF spin fluctuations by antinodal charges, which leads to 
the opening of the pseudogap.
(b) and (c) illustrate that the spin and charge soft modes, respectively,
can be viewed as Yakawa-type bonding bosons, with the ``vacuum'' corresponding
to the situation having dynamic stripe spin-charge modulations.
The energy and momentum input by neutrons to nodal charges liberate an AF spin fluctuation
to this ``vacuum equivalent'', leading to the resonance mode, while
energy input of ARPES photon liberates the charge fluctuations
(by breaking the charge-density wave pair formed by antinodal charges),
leading to the ARPES SCP. 
These liberation processes 
occur at a cost of condensation energy,
corresponding to the twin spin/charge roton energy, with
the intensities proportional to the superfluid density $n_{s}/m^{*}$.}
\label{fig11}
\end{figure}
%%\end{figure}

Figure 10 shows a motion of a hole in dynamic antiferromagnetic (AF) correlations
of surrounding Cu moments.  When the AF pattern is static, a charge motion 
towards the Cu-O bond direction (with the antinodal momentum) would cause 
three frustrated bonds (Fig. 10(a)).  However, when this charge motion occurs in sequence
with half a period of AF fluctuations, frustrations can be avoided (Fig. 10(b)).
Within one period, the charge would proceed twice the lattice constant $a$.
This process brings a very strong coupling between the AF fluctuations with  
$\hbar\omega_{AF}$ and the antinodal charges at the zone
boundary with $k = 2\pi/2a$ having
the Fermi energy $kT_{F} = \hbar\omega_{AF}$.  
Only when accompanied by a resonant AF fluctuation
can an antinodal charge move freely in the system.

Thus, antinodal charge motion should always be associated with
an AF spin fluctuation, similarly to the charge motion in BCS
superconductors associated with lattice deformation, i.e., a phonon. 
The existence of a AF fluctuation can provide an energy benefit to 
the second antinodal charge.  This results in a ``scattering'' process
which couples antinodal charges with opposite momentum, as shown
in Fig. 11(a).  Unlike BCS, however, this coupling is not retarded.   
This process would create a substantial attractive interaction
among antinodal charges, which would lead to the opening of the 
``pseudo gap'' [1].  

As the energy scale $kT_{F}$ is much higher than the mode energy
of roton-analogue spin-charge modulations, the AF fluctuations
in the above mentioned process are mostly high energy fluctuations near ($\pi,\pi$)
well above the soft-mode energy.
%illustrated by the shaded region in Fig. 11 .
The resonant process would couple such high energy spin fluctuations
with the partner charge fluctuation/motion near the zone center,
having comparable energies.
Increasing charge doping in HTSC would increase this charge
energy scale $kT_{F}$.  When $kT_{F}$ exceeds the spin energy scale $J$,
as presumably happening in the overdoped region, the resonant coupling
mechanism is lost.  Charge doping in cuprates also weakens spin correlations
via introduction of frustration.  These features explain the loss of pairing,
followed by emergence of ``phantom superconductivity'' in the overdoped region.   

\section {Difference between nodal and antinodal charges}

In contrast to antinodal charges, nodal charges are moving on the 
same spin sublattice, as shown in Fig. 10(d).  Therefore, nodal
charges are not subject to spin frustration, regardless of how
their energies compare with energies of AF fluctuations.
This fundamental difference between the ``spin-sensitive'' antinodal
and ``spin-insensitive'' nodal charges would explain 
why the ARPES Fermi-energy spectral weight [26] resides mainly in the nodal 
region in the underdoped cuprates having strong AF fluctuations.
The antinodal charges dominate ARPES response in the overdoped 
region where AF correlations die away, and where the antinodal
charges become ``spin-insensitive''.

If one tries to show the spin and charge roton-like modes in 
a single-particle diagram of Fig. 11(b) and (c), the spin mode appears
as connecting two nodal charges and the charge mode
connecting two antinodal charges, analogous to 
nucleon binding via a Yukawa meson.  In 
these diagrams, ``vacuum'' corresponds to the situation
with dynamic spin/charge stripe correlations.
Neutron scattering then
liberates the S=1 AF magnon, while ARPES liberates the charge
fluctuation, into such a ``vacuum equivalent'', with the energy
input corresponding to the condensation energy, as shown
in Fig. 11(b) and (c).  The intensity of these liberation processes should be proportional
to the superfluid density, as was actually observed in neutron 
and ARPES measurements [16,23].

\section {Phase fluctuations, roton-like excitations, and Nernst region}

In purely 2-d systems, such as a thin film of superfluid He on regular or porous
media, superfluidity is destroyed via liberation of vortex-antivortex pair
at the Kosterlitz-Thouless (KT) transition temperature $T_{KT}$, where the 
superfluid density $n_{s}/m^{*}$ at $T=T_{KT}$ follows the system-independent
universal value given by the KT theory [27].  Since the roton energy scale is much higher
than $T_{KT}$, the 2-d superfluid density in typical
He films shows little temperature dependence
at $0 < T < T_{KT}$, as shown in Fig. 5(a) of ref. [1].  In He thin-films, ``dynamic 
superfluidity'' with dissociated vortices exists at $T_{KT} < T < T_{BE}$ = 3.2 K,
as can be seen in the specific heat results.     

In cuprate systems, with increasing temperature, the superfluid density is 
destroyed by the nodal pair-breaking excitations at low temperatures
up to $T \sim T_{c}/2$, presumably followed by the reduction due primarily 
to excitations of the 
twin spin-charge soft mode at $T_{c}/2 < T < T_{c}$. For the case
of highly 2-d cuprates, such as Bi2212, the system undergoes the KT
transition with sufficiently reduced
superfluid density at $T_{c}$ [28].  In systems close to the magnetic states,
such as the 214 systems near the 1/8 doping concentration,
$T_{c}$ is determined by the soft mode energy, i.e., the closeness
to the competing state.  In either types of cuprates,
the normal state above $T_{c}$ exhibits ``dynamic superconductivity'', as 
was demonstrated by the observation of 
Nernst effect [29] shown in Fig. 12.

\begin{figure}[b]
\centering
% uncomment following line when using a real figure
\includegraphics[width=2.5 in]{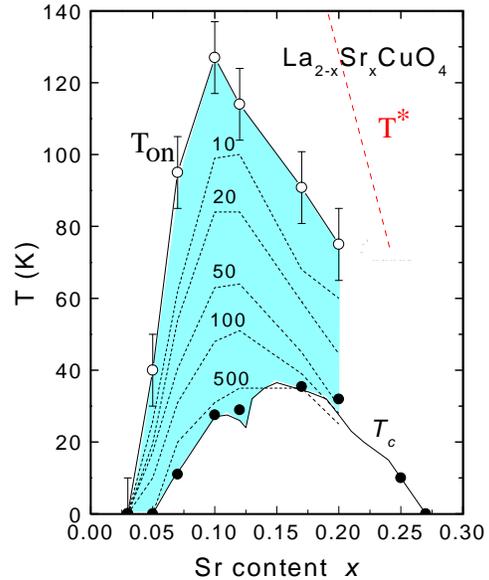}
\caption
{(color)
The region of the Nernst effect, shown in the $T$-$x$ phase diagram
for La$_{2-x}$Sr$_{x}$CuO$_{4}$ (LSCO) [29].  The on-set
temperature of the Nernst effect is denoted by $T_{on}$.
The result of $T_{on}$ for the $x=0.10$ sample
is plotted with the green star symbol in Fig. 2.
The value of $T^{*}$ was taken from [30].} 
\label{fig12}
\end{figure}

For bosonic systems where the pairing energy scale $T^{*}$ is higher than
the number-density scale $T_{B}$, we expect the Nernst region to 
extend up to $T_{B}$.  Indeed, the Nernst
on-set temperature $T_{on} \sim$ 130 K of 10\%\ Sr doped LSCO system [29]
(green star)  
lies very close the the $T_{B}$ line in Fig. 2 [1], and $T_{on}$
increases with increasing doping in the very underdoped region    
in Fig. 12, following the behavior of $T_{B}$.  
The reduction of $T_{on}$ near the optimum doping region
in Fig. 12 can be explained by the reduction of $T^{*}$,
which makes bosons unavailable.  Thus, the Nernst effect
occurs with two conditions:  (1) availability of bosons, i.e., $T < T^{*}$; 
and (2) sufficient number density of bosons to build up local phase coherence, 
i.e. $T < T_{B}$.  The Nernst region above $T_{c}$ can be characterized by the 
existence of bosonic amplitude without static long-range
phase coherence.  If it were possible to remove 
factors which destroy long-range phase coherence, such as the soft mode and low dimensionality,
cuprate systems would have acquired superconductivity at the Nernst onset temperature
$T_{on}$.

Since the soft-mode energy is related to condensation energy,
superfluid density influences the mode energy.  However,
due to relevance of competing states,
$n_{s}/m^{*}$ at $T\rightarrow 0$ is no longer a sole factor in determining $T_{c}$.  This feature
is clearly seen for the case of 214 cuprates, which are
particularly close to the competing magnetic state, 
having lower soft-mode energies, thus showing earlier
branching off from the linear relationship in Fig. 1,
compared to YBCO systems with comparable values of 
$\sigma(T\rightarrow 0) \propto n_{s}/m^{*}$.
In this way, superfluid density and the soft-mode energy
``conspire or cooperate'' in determining $T_{c}$. 
Further studies of microscopic pairing mechanisms and  
collective modes would hopefully clarify the origin of 
apparently universal upperlimit of $T_{c}/T_{B} \sim$ 1/4
common to various correlated electron superconductors 
shown in Fig. 2.\\

%\section{Acknowledgement}
This work has been supported by the NSF 
DMR-0102752, DMR-0502706, INT-0314058 and CHE-0111752 (Nanoscale Science and Engineering
Initiative).  The author thanks J.C. Davis, A. Damascelli, N. Nagaosa for useful discussions;
G.M. Luke and S. Uchida for continuing collaboration; and T. Yamazaki for excellent
guidance in the beginning of the author's scientific career.

\end{document}